\documentclass[12pt]{article}

\begin{document}
\thispagestyle{empty}
\noindent\
\\
\\
\\
\begin{center}
\large \bf The Size of the Weak Bosons

\end{center}
\hfill
 \vspace*{1cm}
\noindent
\begin{center}
{\bf Harald Fritzsch}
\vspace*{0.5cm}\\

University of Munich, Germany\\
Faculty of Physics\\
Arnold Sommerfeld Center for Theoretical Physics

and\\

Nanyang Technological University, Singapore\\ 
Institute for Advanced Study
\vspace*{0.5cm}\\
\end{center}

\hfill\\
\hfill\\

\begin{abstract}
We study the hypothesis that the bosons are composite systems, which have a size of the order of $10^{-17}$ cm. The electromagnetic selfenergies of the weak bosons lead to specific departures from the standard elektroweak model, in agreement with observation. Above the energy of 1 TeV the standard electroweak
model breaks down completely. \\
\end{abstract}

\newpage

The weak bosons have a large mass, unlike the massless gauge bosons (photons, gluons). For this reason they might not be elementary gauge bosons, but composite particles, consisting of an elementary fermion and its antiparticle. They would be similar to the $\rho$-mesons in $QCD$, which are bound states of a quark and an antiquark. \\

The size of the $\rho$-mesons, which is of the order of $10^{-14}$ cm, is determined by the scale parameter of quantum chromodynamics $\Lambda_c$, which has been measured in the experiments: $\Lambda_c$ =  $217 \pm 25$ MeV. The inverse size of the weak bosons must be at least thousand times larger than $\Lambda_c$.\\

In this paper I study the hypothesis that the weak bosons are composite systems (see also ref.(1)). I estimate the size of the weak bosons, using the present experimental data, in particular the electromagnetic self energies. In a composite model of the weak bosons the electromagnetic self energies contribute to the masses of the weak bosons, however these contributions are very similar to the radiative corrections in the standard model due to the virtual top quark and due to the "Higgs" boson.\\

According to the experiments the mass difference between the charged and neutral $\rho$-meson must be very small, not larger than 0.7  MeV (ref. (2)). Both the masses of the charged and of the neutral $\rho$ -mesons change, when the electromagnetic interaction is introduced. The mass of the charged $\rho$-meson increases due to the electromagnetic interaction of the quarks, the mass of the neutral  $\rho$-meson decreases.\\

Thus the mass difference should be about the same as for the pions, about 4.6 MeV, in disagreement 
with observation.  However there is also a mixing of the neutral $\rho$-meson and the photon, which increases the mass of the neutral  $\rho$ -meson.  The mass shift due to mixing can be calculated. It depends on a mixing parameter $\mu$, which is determined by the electric charge, the decay constant $F_\rho$ and the mass of the $\rho$-meson:\\

\begin{equation}
\mu \; =\; e \frac{F^{}_\rho}{M^{}_\rho}~ .
\end{equation}\\

The mass shift due to the mixing is given by:\\

\begin{equation}
M^2_{\rho^0} - M^2_{\rho^+} \; =\; M^2_{\rho^+} \left(\frac{\mu^2}{1 - \mu^2}\right) \; .
\end{equation}\\

The decay constant is measured to about 220 MeV, which gives $\mu\approx 0.09$. It leads to a mass shift of about 3 MeV. Due to this mass shift the mass difference between the charged and neutral $\rho$-meson is very small. The mixing increases the mass, but the electromagnetic interaction between quark and antiquark decreases it. Both effects nearly cancel each other. \\

We assume that the weak bosons consist of a basic lefthanded fermion and its antiparticle, which are denoted as "haplons" - the Greek translation of "simple" is "haplos".  A theory of this type was proposed in 1981 (see  ref.(3), also ref.(4,5,6,7,8)). The new confining gauge theory is denoted as quantum haplodynamics ($QHD$). The $QHD$ mass scale is given by a mass parameter $\Lambda_h$, which determines the size of the weak bosons. \\

Two types of  haplons are needed as constituents of the weak bosons, denoted by $\alpha$ and $\beta$.  Their electric charges in units of e are:\\

\begin{equation}
h = \left( \begin{array}{l}
+\frac{1}{2}\\
-\frac{1}{2}\\
\end{array} \right) \ .
\end{equation}\\

The three weak bosons have the following internal structure:\\
\begin{eqnarray}
W^+ & = & \overline{\beta} \alpha \; , \nonumber \\
W^- & = & \overline{\alpha} \beta \; , \nonumber \\
W^3 & = & \frac{1}{\sqrt{2}} \left( \overline{\alpha} \alpha -
\overline{\beta} \beta \right) \; .
\end{eqnarray}\\

The weak bosons are degenerate in mass in the absence of electromagnetism. If the electromagnetic interaction is introduced,
the mass of the neutral boson increases due to the mixing with the photon. In addition the masses are changed due to the electromagnetic self energies of the weak bosons. The electromagnetic interaction leads to a reduction of the mass of the neutral weak boson and to an increase of the mass of the charged weak boson. These electromagnetic self energies are not present in the standard electroweak model. They lead to definite departures from the standard model. Similar departures are also expected in the standard electroweak model, caused by the radiative corrections.\\

The mixing of the weak bosons is described by a mixing parameter m, which is determined by the decay constant of the weak boson $F_W$, defined in analogy to the decay constant of the $\rho$-meson in $QCD$ (ref.(5)):\\

\begin{equation}
\langle 0 \left| \frac{1}{2} \left( \overline{\alpha} \gamma^{}_\mu
\alpha - \overline{\beta} \gamma^{}_\mu \beta \right) \right| Z
\rangle \; =\; \varepsilon^{}_\mu M^{}_W F^{}_W \; .
\end{equation} \\

\begin{equation}
m \; =\; e \frac{F^{}_W}{M^{}_W} \ .
\end{equation} \\

In the standard electroweak model the mixing parameter m is given by the weak mixing angle (ref.(5)):\\

\begin{equation}
\sin\theta^{}_w \; =\; m \ .
\end{equation}\\

The mass difference between the $Z$-boson  and the $W$-boson is determined by the mixing parameter m and the $W$-mass (see also eq.(2)): \\

\begin{equation}
M^2_Z - M^2_W \; =\; M^2_W \left(\frac{m^2}{1 - m^2}\right) \; .
\end{equation}\\

We shall use the following experimental values:\\
\begin{eqnarray}
M^{}_W & = & 80.384 \pm 0.014 ~{\rm GeV} \; , \nonumber \\
M^{}_Z & = & 91.1874 \pm 0.0021
 ~{\rm GeV} \; , \nonumber \\
F^{}_W & \approx \ & 123.9 ~{\rm GeV} \; , \nonumber \\
\sin^2 \theta^{}_W & = & 0.2315 \; , \nonumber \\
\alpha & = & \frac{e^2}{4\pi} \; \approx \; \frac{1}{128.9} \; .
\nonumber \\
\end{eqnarray} \\

In the standard electroweak model one can calculate the mass of the charged weak boson, taking into account the mass of the Z-boson and the 
electroweak mixing angle. The result includes radiative corrections, which in particular depend on the mass of the t-quark and 
the unknown mass of the Higgs boson.\\

In our approach the mass difference between the charged and neutral weak bosons depends also on the electromagnetic self energies of the weak bosons. The masses of the weak bosons are given by:\\

\begin{eqnarray}
M(W) = M + \Delta M \; \nonumber \\
M(Z) = \frac{M}{\sqrt{1 - m^2}} - \Delta M \; . \nonumber\\  
\end{eqnarray} \\

Here M is the value of the charged weak boson mass in the absence of the electromagnetic interaction, and $\Delta M$ is the electromagnetic self energy, which is positive for the charged weak boson and negative for the neutral one.\\

We calculate M and $\Delta M$ for particular values of the mixing parameter m:\\

\begin{eqnarray}
m^{2} =0.2300  ~  ~  ~ => ~  M = 80.188 ~{\rm GeV} ~  ~ || ~ ~   \Delta M = 0.196 ~ GeV \;  \nonumber \\
m^{2} =0.2350  ~  ~  ~ => ~  M = 80.049 ~{\rm GeV} ~  ~ || ~ ~   \Delta M = 0.335 ~ GeV \;  \nonumber \\
m^{2} =0.2400  ~  ~  ~ => ~  M = 79.909 ~{\rm GeV} ~  ~ || ~ ~   \Delta M = 0.475 ~ GeV \; . \nonumber \\
\end{eqnarray} \\

For the central value $m^2$=0.2315 we find M = 80.147 GeV and $\Delta M$ = 0.237  GeV. The range of values for $m^2$, considered above, covers the range, given by the present experimental values.\\

The mass $\Delta M$ is the electromagnetic selfenergy of the charged weak boson. It is given by the product of the electric charges of the two haplons inside the weak boson, the finestructure constant, the $QHD$-scale $\Lambda_{h}$ and a constant c:\\

\begin{equation}
\Delta M \approx\frac{1}{4}~{\alpha} ~c~\Lambda_{h}~ 
\end{equation}\\

>From this equation we can determine the value of $\Lambda_{h}$: \\

\begin{equation}
\Lambda_{h}\approx 516~ \Delta M/c ~.
\end{equation}\\

The constant c is expected to be of the order of one. The precise value depends on dynamical details of $QHD$ and cannot be calculated. We obtain for the $QHD$ mass scale, taking into account eq.(11):\\  

\begin{equation}
100 ~  GeV <c \Lambda_{h} <  250 ~ GeV~ .
\end{equation}\\

The radius of a weak boson will depend on the precise values of c. If we take c=1, the radius will be in the range $0.001  - 0.002~
 fm$.\\

The  precise value of the $QHD$ mass scale depends on the parameter c. It should be between 0.05 TeV and 0.5 TeV. It is about thousand times larger than the  $QCD$ mass scale.\\ 

In strong interaction physics above the energy of 1 GeV many resonances exist. We expect similar effects above 1 TeV in the electroweak sector. Above 1 TeV there should in particular exist excited weak bosons, which will decay mainly into two or three weak bosons. These states could be observed soon at the $LHC$.\\

We conclude: If our hypothesis is correct, the size of the weak bosons should be of the order of $0.002~ fm$. In this case the internal structure of the weak bosons should soon be discovered at the Large Hadronic Collider at CERN.\\

{\end{document} }}